\documentstyle[aps,12pt,manuscript]{revtex}
\begin{document}
\def\overlay#1#2{\setbox0=\hbox{#1}\setbox1=\hbox to \wd0{\hss #2\hss}#1%
\hskip -2\wd0\copy1}
\preprint{ INJE-TP-98-10, hep-th/9809172}

\title{Entropy Problem in the  AdS$_3$/CFT correspondence}
\author{ Y. S. Myung }
\address{Department of Physics, Inje University, Kimhae 621-749, Korea} 

\maketitle

\vskip 2in

\begin{abstract}
We resolve the entropy problem in the AdS$_3$/CFT correspondence by introducing
both the normalizable  and non-normalizable bulk modes.
On the boundary, the normalizable Liouville states gives us $c=1$ conformal field theory(CFT),
whereas the non-normalizable Liouville states provide $c = {3 \ell \over 2 G}$ CFT.
Such (boundary) non-normalizable modes come from non-normalizable bulk modes
which serve as classical, non-fluctuating bulk background and
encode the choice of local operator insertion on the boundary.
Since the non-normalizable bulk modes can transfer information from bulk to boundary, 
it suggests that counting of non-normalizable states on the boundary at infinity leads to the 
entropy($S={2 \pi r_+ \over 4 G}$) of (2+1)-dimensional gravity with $ \Lambda= -1/\ell^2$.

\end{abstract}

\newpage

Recently the AdS/CFT correspondence has attracted much interest. This is based on the duality
relation between the string theory(bulk theory) on AdS$_{d+1}$ and a conformal field theory(CFT)
as its d-dimensional boundary theory\cite{Mal2,Gubs,Witt}. 
The calculation of greybody factor for the BTZ black hole on AdS$_3$ 
is in agreement with the CFT calculation
on the boundary\cite{Birm,Lee1,Lee2}. Here one introduced a set of test fields
to connect the bulk to the boundary. 
These couple to conformal operators on the boundary : the tachyon couples to (1/2,1/2) operator,
while the dilaton couples to (2,2) operator.
Actually we calculated the greybody factor of test fields by using their non-normalizable
modes and AdS$_3$/CFT correspondence.
The normalizable and non-normalizable modes emerge natually  from a direct solution
of the wave equation($\nabla^2 \Phi - m^2 \Phi =0$) in the BTZ background($r,t,\varphi$) as\cite{Lee2,Vak}

\begin{equation}
\Phi = \Phi^{nor} + \Phi^{non-nor} = c_1 x^{-( 1 + \sqrt{1+ m^2\ell^2})}
         + c_2 x^{( -1 + \sqrt{1+ m^2\ell^2})}
\end{equation}
when $x=r/\ell$ is large. Here $m^2\ell^2=(h+\bar h)(h+ \bar h-2)$ and $c_1,c_2$ are constants.
The first (second) terms with $m^2\ell^2 > 0$ converges (diverges) at $x=\infty$
and thus turn out to be the normalizable(non-normalizable) modes.
This means that $\Phi^{nor}$ satisfies the Dirichlet boundary condition at spatial infinity,
but $\Phi^{non-nor}$ does not. In the case of $m^2\ell^2=8$(dilaton), we find both 
the normalizable(${ 1 \over x^4}$) and non-normalizable($x^2$) modes.
For $m^2\ell^2=0$(a free scalar) one has a constant at $x=\infty$ and for $m^2\ell^2=-1$
(tachyon) one has only normalizable modes.
$\Phi^{nor}$ propagates in the bulk and corresponds to physical state, while $\Phi^{non-nor}$
plays a role of classical, non-fluctuating background. We emphasize that the non-normalizable
modes encode the local operator insertions on the  boundary and thus correspond to
introducing the particular boundary condition. Hence, on the basis of $\Phi^{non-nor}$,
 one can calculate
the greybody factor of relevant fields\cite{Lee1,Lee2}. The result is consistent with that of dual CFT
on the boundary\cite{Gub}.
This proves  the $AdS_3$/CFT correspondence concerning the greybody factor(dynamical property).
Here the non-normalizable modes plays a role of messenger to transfer  information 
from bulk to  boundary.

On the other hand, nowdays it seems to be a discrepancy for counting the entropy
(static property) of 
the BTZ black hole in relation to the AdS/CFT correspondence.
Here two important questions arise : 1) what fields provide the relevant degrees of freedom?
 2) where do these excitations live?
An answer is that the freedom at infinity is relevant to counting of the entropy.
We agree that the freedom at infinity is given by
a single Liouville field. But there are two central charges ($c=1$\cite{Kut,Cous,Calip,Mart2} and 
$c = {3 \ell \over 2 G}$\cite{Brow,Seib,Stromi,Bana,Behr,Give}) for the Liouville model. 
The problem is which one describes the boundary states correctly.
Hence we have to understand a discrepancy between $c=1$ and  $c= {3 \ell \over 2G}$.

In this paper, we  wish to resolve this urgent problem.
Aside from test of three and four point functions\cite{Muk,Liu,Fre}, the correct counting of entropy
in both bulk and boundary will prove the validity of AdS$_3$/CFT correspondence.

On the dual CFT side, one finds 
the asymptotic density of states at infinity(by Cardy)

\begin{equation}
\ln \rho(\Delta,\bar \Delta) \sim 2 \pi \sqrt{ c_R \Delta \over 6} + 
                  2 \pi\sqrt{ c_L \bar \Delta \over 6}
\end{equation}
with  central charges
$c_L=c_R= {3 \ell \over 2G}$ for two copies of Virasoro algebra\cite{Brow,Stromi}.
Here $\Delta,\bar \Delta$ are the eigenvalues of two Virasoro generators $L_0, \bar L_0$.
Considering the BTZ black hole with mass $M=(L_0 +\bar L_0)/\ell$ and angular momentum
$J= L_0 -\bar L_0$,
we obtain the correct Bekenstein-Hawking entropy
\begin{equation}
S_{B-H}={2 \pi r_+ \over 4 G}.
\end{equation}

Here we introduce a  set of test fields($\{\Phi_i \}$) with normalizable
and non-normalizable modes to resolve a discrepancy.
 This corresponds to introducing
a Liouville field with normalizable and non-normalizable modes on the boundary.
First we wish to discuss the normalizable modes.
These are quantized in the bulk and correspond to states in the boundary Hilbert space. 
In other words, the  normalizable  bulk modes  can be realized as normalizable
 Liouville modes on the  boundary,
whose lowest Virasoro eigenvalue is given by $\Delta=(c_{Liou}-1)/24$.
How can we  realize normalizable bulk modes as normalizable
 Liouville modes on the  boundary?
This can be understood from a mechanism of conformal anomaly inflow onto the 
boundary without non-normalizable modes\cite{Mart1}.
 Thus, instead of $c_R$ in Eq.(2), one has to use the effective central charge
($c_{eff}= c_{Liou} -24 \Delta =1$)\cite{Calip}.
Recently this mechanism  can be further clarified  by Martinec\cite{Mart2}.
He showed that (2+1)-dimensional gravity with $\Lambda=-1/\ell^2$ is a pure gauge theory
and thus it examines  only its macroscopic properties(thermodynamics) by a set of 
Noether charges. 
Hence the boundary theory of (2+1)-dimensional gravity appears as
 a collective field excitation of the microscopic
dual CFT. Its effective bulk/boundary theory turns out to be 
$SL(2,{\bf R})_L \times SL(2,{\bf R})_R$ Chern-Simons/$c=1$ Liouville
theory. Also this can be confirmed by analogy with  quantum Hall effect, whose effective theory
is $U(1)$ Chern-Simons/$c=1$ chiral boson\cite{myung}.
The boundary theory of (2+1) gravity with $\Lambda=-1/\ell^2$($c=1$ normalizable
Liouville modes) cannot
be used for the correct counting of  microscopic states of the BTZ black hole.

Now we are in a position to remark the non-normalizable modes.
In order to construct the microscopic
structure on the boundary, we need a set of  test fields ($\{\Phi_i\}$) with non-normalizable modes.
These can be used for the insertion of conformal operators on the boundary and thus 
make a microscopic structure on the boundary.
In turn, these non-normalizable modes can be expressed  
by  non-normalizable Liouville modes with $\Delta=0$\cite{Calip}.
This is so because the Liouville model can be considered as a $\sigma$-model description
for a set of test fields(for example, tachyon($T$), dilaton($\phi$), $\cdots$)
\begin{equation}
S_\sigma = {k-2 \over 4 \pi} \int_{\partial M_\infty} 
\big (  \partial_u \lambda \partial_v \lambda 
     +   T(\lambda) + Q R^{(2)} \phi(\lambda)  \big ),
\end{equation}
where  $k=\ell/4G$, and $R^{(2)}$ is the Ricci curvature of boundary world sheet
($u=t-\varphi,v=t+\varphi$) at infinity\cite{Behr}.
The Liouville field($\lambda=\ln x$) corresponds to the radial coordinate of the
target space.
 As a result, the non-normalizable Liouville
modes
correspond to operator insertions and thus account for the microscopic structure 
of the boundary CFT.
On the other hand, we remind the reader that
 the gauge theory of branes (dual CFT) is a tool to investigate
its  microscopic features. And thus we have to investigate operator insertions
to count the boundary states rather than considering the Hilbert space of normalizable
modes. Here instead of  operator insertions(non-normalizable bulk modes), one can use Liouville
modes to obtain the central charge. In this case one finds it by calculating the dilaton$\beta$-
function as 
\begin{equation}
\tilde c_{Liou}= 1 + 6(k-2)Q^2= {3k \over k-2} -2 +6k,
\end{equation}
where $Q$ is a background coupling charge($Q=(1-k)/(k-2)$).
It is worth noting that the tachyon contains only  normalizable modes.
From Eqs.(4) and (5), if the dilaton is absent, $\tilde c_{Liou}$ leads to 1\cite{Cous}.
This is just the central charge for normalizable Liouville modes.
It accounts for  normalizable tachyonic bulk modes.
If the dilaton is present, one can lead to  $c = {3 \ell \over 2 G}$ CFT.
This is so because the dilaton has non-normalizable bulk modes 
and thus couples to (2,2) operator on the boundary.
On the boundary the Liouville modes represent  these non-normalizable modes.

The central charge $\tilde c_{Liou}$ 
is determined by the coupling constants and can be chosen  to be arbitrarily large.
In the classical limit of $k\to\infty$, the last term($6k$) is relevant and this 
leads to $c = {3 \ell \over 2 G}$.

In conclusion, $c=1$ Liouville theory is the effective boundary theory for
normalizable bulk modes, whereas $c = {3 \ell \over 2 G}$ Liouville model is the 
effective  boundary description for non-normalizable bulk modes.
Since the non-normalizable modes connect the bulk information to boundary correctly,
the relevant one to the AdS$_3$/CFT correspondence is just 
$c = {3 \ell \over 2 G}$ Liouville model. Then one finds the same entropy as in Eq.(3)
by the boundary theory. This confirms the  AdS$_3$/CFT correspondence in relation to
the entropy of the BTZ black hole.
Finally it is noted that we introduce only two test fields(tachyon and dilaton).
In order to understand this picture fully, one needs a complete set of test fields.

\acknowledgments
I would like to thank J. Maldacena, E. Martinec and H. W. Lee for helpful discussions.
This work was supported in part by the Basic Science Research Institute 
Program, Ministry of Education, Project NO. BSRI--98--2413 and Grant from Inje University, 1998.

\end{document}